\documentclass[runningheads]{svmult}

\usepackage{graphicx}  
\usepackage{multicol}  
\usepackage{physprbb}  

\def\gtrsim{\;\rlap{\lower 2.5pt
 \hbox{$\sim$}}\raise 1.5pt\hbox{$>$}\;}
\def\lesssim{\;\rlap{\lower 2.5pt
   \hbox{$\sim$}}\raise 1.5pt\hbox{$<$}\;}

\begin{document}

\title*{How black holes get their kicks:\protect\newline Radiation
recoil in binary black hole mergers}

\toctitle{How black holes get their kicks:\protect\newline Radiation
recoil in binary black hole mergers}

\titlerunning{Radiation recoil in binary black hole mergers}

\author{Scott A.\ Hughes\inst{1}
\and Marc Favata\inst{2}
\and Daniel E.\ Holz\inst{3}}

\authorrunning{Scott A.\ Hughes et al.}

\institute{Department of Physics and Center for Space Research,
Massachussets Institute of Technology, 77 Massachusetts Avenue,
Cambridge, MA 02139
\and Department of Astronomy, Cornell University, 611 Space Sciences
Building, Ithaca, NY 14853
\and Center for Cosmological Physics, University of Chicago, Chicago,
IL 60637}

\maketitle

\begin{abstract}
Gravitational waves from the coalescence of binary black holes carry
linear momentum, causing center of mass recoil.  This ``radiation
rocket'' has important implications for systems with escape speeds of
order the recoil velocity.  We describe new recoil calculations using
high precision black hole perturbation theory to estimate the
magnitude of the recoil for the slow ``inspiral'' coalescence phase;
coupled with a cruder calculation for the final ``plunge'', we
estimate the total recoil imparted to a merged black hole.  We find
that velocities of many tens to a few hundred km/sec can be achieved
fairly easily.  The recoil probably never exceeds about 500 km/sec.
\end{abstract}

It is very well known that gravitational waves (GWs) carry energy and
angular momentum from a binary system, causing decay of the binary's
orbit and eventually driving the system to merge into a single object.
Although it has been understood for quite some time (e.g.,
{\cite{b73}}), it is somewhat less well-appreciated that these waves
can carry {\it linear} momentum from the system as well.  The center
of mass in this case must recoil in order to enforce global
conservation of momentum.  If the recoil velocity is comparable to or
greater than the escape velocity of the binary's host structure, there
could be important dynamical consequences, such as ejection of the
merged black hole remnant.

The recoil arises because the radiation field generated by a binary is
typically asymmetric.  As a helpful cartoon, consider the following
argument due to Alan Wiseman.  In an unequal mass binary (Fig.\
{\ref{fig:schematic}), the smaller member, $m_1$, moves with a higher
speed than the larger member, $m_2$.  It is thus more effective at
``forward beaming'' its wave pattern.  This means that there is an
instantaneous net flux of momentum ejected from the system parallel to
the velocity of the smaller body, and a concomitant recoil opposing
this.

\begin{figure}[t]
\begin{center}
  \includegraphics[width=.5\textwidth]{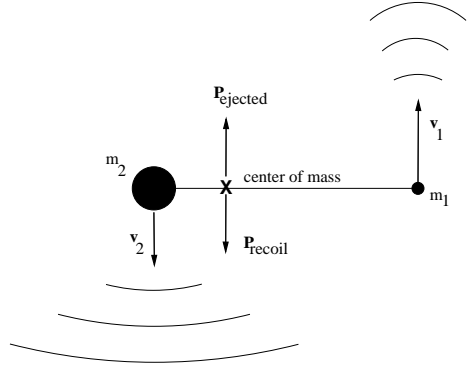}
\end{center}
\caption{GW emission from an unequal mass binary.  Momentum is ejected
parallel to the smaller body's velocity (${\vec v}_1$).  Conservation
of momentum requires that the system recoil in the opposite
direction.}
\label{fig:schematic}
\vskip -4mm
\end{figure}

Over an orbit, the recoil direction continually changes.  If the orbit
were perfectly circular, this means that there would be no net
interesting effect --- the binary's center of mass would run around in
a circle, and the {\it net} recoil would sum to zero.  However, when
GW emission is strong, the orbit is {\it not} perfectly circular:
Because of the secular, dissipative evolution of the binary's energy
and angular momentum, the black holes slowly spiral towards one
another.  Since the orbit does not close, the recoil does not sum to
zero.  The recoil accumulates until the holes merge and settle down to
a quiescent state, shutting off the momentum flux and yielding a net,
non-zero kick.

This recoil is not a wierd property of GWs --- it holds for {\it any}
form of radiation\footnote{Indeed, electromagnetic or neutrino recoil
may impact neutron star kicks {\cite{ht75,lcc01}}.}.  This can be
brought out by considering a multipolar decomposition.  Suppose we
build a distribution of charges that has a non-zero electric dipole
and quadrupole moment, as in Fig.\ {\ref{fig:dipquad}}.  Suppose
further that we spin this charge arrangement about its center point,
driving the system to radiate electromagnetic waves.  What does this
radiation distribution look like from far away?

\begin{figure}
\begin{center}
  \includegraphics[width=.35\textwidth]{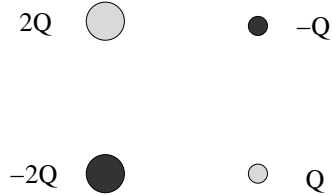}
\end{center}
\caption{Charge distribution with non-zero dipole and quadrupole
moment.  Spinning this distribution about its center point produces
radiation carrying non-zero linear momentum due to beating between the
dipolar and quadrupolar radiation fields.}
\label{fig:dipquad}
\vskip -5mm
\end{figure}

The radiation's {\it amplitude} has two pieces, dipole and quadrupole:
\begin{eqnarray}
{\vec E} &=& {\vec E}^{\rm dip} + {\vec E}^{\rm quad}\;,\qquad {\rm where}
\\
{\vec E}^{\rm dip} &\propto& e^{i(\phi - \omega t)}\;,\qquad
{\vec E}^{\rm quad} \propto e^{2i(\phi - \omega t)}\;.
\end{eqnarray}
Since the intensity $I \propto |\vec E|^2$, it will contain three
pieces:
\begin{equation}
I = I^{\rm dip} + I^{\rm quad} + I^{\rm dip-quad}\;,
\end{equation}
where
\begin{eqnarray}
I^{\rm dip} &\propto& |\vec E^{\rm dip}|^2 \propto {\rm constant}\;;
\qquad
I^{\rm quad} \propto |\vec E^{\rm quad}|^2 \propto {\rm constant}
\\
I^{\rm dip-quad} &\propto& {\rm Re}\left[\vec E^{\rm dip}{\bar{\vec
E}}^{\rm quad}\right] \propto \cos(\phi - \omega t)\;.
\end{eqnarray}
The intensity has a preferred direction, which rotates as the charge
distribution rotates.  The energy from the system is instantaneously
beamed in a preferred direction, and so there is a net flux of
momentum in that direction as well.

Since the lowest order GWs are quadrupolar, recoil from GW emission
must come (at lowest order) from a beating of the mass quadrupole with
mass octupole and current quadrupole moments.  The mass octupole and
current quadrupole vanish for an equal mass binary, demonstrating ---
in accord with our ``forward beaming'' intuition --- that unequal
masses are needed for there to be any recoil.  This also demonstrates
that GW recoil must be a very small effect, except perhaps in the very
late stages of coalescence --- the octupole radiation amplitude is
smaller than the quadrupole by a factor of order $v/c$ (where $v$ is
orbital speed).

The first careful analysis of recoil in binary systems due to GW
emission is that of Michael Fitchett {\cite{f83}}.  Fitchett's
analysis described the orbital dynamics of the binary using Newtonian
gravity and only included the lowest radiative multipoles which
contribute to the recoil.  His analysis predicted that the recoil of
the merged remnant took the form
\begin{equation}
v_{\rm F} \simeq 1480\,{\rm km/sec}\frac{f(q)}{f_{\rm max}}
\left(\frac{2 G(m_1 + m_2)/c^2}{R_{\rm term}}\right)^4\;,
\label{eq:vfitchett}
\end{equation}
where $R_{\rm term}$ is the orbital separation at which GW emission
terminates, $q = m_1/m_2$ is the binary's mass ratio, and $f(q) =
q^2(1 - q)/(1 + q)^5$ is a function whose maximum is at $q \simeq
0.38$, and has the limit $f(q) \simeq q^2$ for $q \ll 1$.

Three features of this formula are particularly noteworthy.  First,
this result does not depend on total mass --- only on the mass {\it
ratio} (bearing in mind that $R_{\rm term}$ scales with total mass
$M$).  Thus, this scaling holds for any binary black hole merger ---
stellar mass mergers through supermassive mergers.  Second, the
overall scale is quite high.  Although there is an important
dependence on mass ratio and the termination radius $R_{\rm term}$ is
somewhat uncertain, Eq.\ (\ref{eq:vfitchett}) indicates that kicks of
hundreds of km/sec are not difficult to achieve; kicks $\gtrsim 1000$
km/sec are plausible.  This is high enough that we might expect black
hole ejection following a merger to be common.

Notice, however, that the recoil becomes very strong when the
separation of the bodies is small.  This is a strong hint that we
cannot take Eq.\ (\ref{eq:vfitchett}) at face value --- the strong
gravity physics neglected by Ref.\ {\cite{f83}} is likely to be very
important.

A few efforts have improved on Fitchett's analysis over the years.
Fitchett and Detweiler {\cite{fd84}} first made a strong field
analysis, treating the binary as a Schwarzschild black hole $M$
orbited by a point mass $\mu$.  The orbiting body's influence can then
be studied using black hole perturbation theory.  Their results
suggested that Eq.\ (\ref{eq:vfitchett}) describes the recoil fairly
well.  Wiseman {\cite{w92}} analysed the recoil with post-Newtonian
theory (roughly, an expansion in $v \sim \sqrt{G M/rc^2}$).  He found
that Fitchett's formula tended to systematically overestimate the
recoil by $\gtrsim 10\%$.  Unfortunately, his results behave somewhat
pathologically for $r \lesssim 9 GM/c^2$ due to ill behavior of the
expansion in the very strong field.

This motivates our analysis of this problem.  Our formal setup is
quite similar to that of Fitchett and Detweiler: We model the binary
as a pointlike body $\mu$ moving on the exact, geodesic orbits of a
Kerr black hole with mass $M$ and spin parameter $a$.  We compute the
GWs emitted from such orbits very accurately using black hole
perturbation theory {\cite{t73,h00,h01}}, and extract the recoil from
the wave pattern {\cite{fhh04}}.  The perturbative approach allows us
to study the dynamics of the binary's spacetime with high accuracy.
Schematically, we treat this spacetime as that of the Kerr black hole
plus a small perturbation:
\begin{equation}
g_{\alpha\beta} = g_{\alpha\beta}^{\rm Kerr}(M,a) +
h_{\alpha\beta}(\mu)\;.
\label{eq:perturb}
\end{equation}
By requiring that this spacetime satisfy the Einstein field equations,
it can be shown {\cite{t73}} that $h_{\alpha\beta}$ is governed by a
wave-like equation.  The wave operator automatically captures the most
important properties of the strong field physics.

This approach is strictly accurate only when $\mu \ll M$ --- the
smaller body cannot significantly distort the spacetime if we want the
exact Kerr orbits to describe our binary.  We believe, though, that we
can extrapolate out of this regime with accuracy good enough for most
astrophysical purposes --- our extrapolation errors are estimated to
be at most several tens of percent, at least up to a mass ratio $q
\simeq 0.3$, and to scale with the squared mass ratio.

We focus upon circular, equatorial orbits of Kerr black holes.
Circularity is surely a good approximation, since eccentricity is
rapidly reduced during coalescence.  The equatorial assumption is not
so good; since the binaries of interest form through captures, we
expect no particular alignment between the spin and orbit.  We are
working to lift this approximation, which requires moderately
substantial modifications to our code; early indications are that the
inclination does not have a very large effect on the recoil, other
than to change the radius at which a transition in the orbital
dynamics occurs.  Results from the prograde and retrograde equatorial
orbits appear to bound the recoil at any spin.

One technical detail of our code is important enough that it requires
some explanation.  Our calculation expands the perturbation as
\begin{equation}
\Psi_4 = \frac{1}{r}\sum_{lm} Z_{lm}S_{lm}(\theta)e^{im(\phi -
\Omega t)}\;.
\end{equation}
$\Psi_4$ represents a curvature perturbation, and serves as a
surrogate for $h_{\alpha\beta}$.  $S_{lm}(\theta)$ is a spheroidal
harmonic (very similar to a spherical harmonic), and $Z_{lm}$ is a
complex number found by solving a certain differential equation
{\cite{t73,h00}}.  The most important aspect of this equation for our
purposes is the presence of the orbital frequency $\Omega$: We assume
that the orbit is well-described by a Fourier expansion.  This is only
true if the orbit is periodic (or nearly so).  This in turn is only
true when the separation $r$ of the black holes is greater than the
radius of the ``last stable orbit'' (LSO), $r_{\rm LSO}(a)$.  (For
Schwarzschild holes, $a = 0$, $r_{\rm LSO} = 6GM/c^2$.)

In the regime $r > r_{\rm LSO}(a)$, dynamically stable orbits exist.
GWs slowly evolve the system through a sequence of stable orbits of
ever decreasing radius.  At any instant in this regime, the orbit is
well modeled as periodic with a well-defined frequency $\Omega$.  The
recoil is then simply related to the coefficients $Z_{lm}$ and to an
overlap integral involving the harmonics $S_{lm}(\theta)$; see Ref.\
{\cite{fhh04}} for details.

As we approach and cross $r_{\rm LSO}(a)$, the orbit becomes a rapid
``plunge'' in which the small body quickly falls into the event
horizon.  This final plunge {\it cannot} be treated using this Fourier
expansion, so our calculation breaks down.  We are currently
developing techniques to model this regime accurately.  For the
present analysis, we compute a probable overestimate (based on an
extrapolation of the inspiral momentum flux beyond its range of
validity) and a probable underestimate (based on a low-order flux
formula coupled to the plunging motion) in order to provide a
reasonable range for the likely recoil.

Our analysis shows that Fitchett's calculation {\it consistently
overestimates} the recoil velocity, especially as the LSO is
approached; see the left-hand panel of Fig.\ {\ref{fig:results}}.
This is due to strong-field physics: A wave packet released near the
horizon redshifts as it propagates to large radius, reducing the
energy and momentum that it carries.  This effect is not present in
calculations which neglect curved spacetime physics, as in Ref.\
{\cite{f83}}.  Also, when radiation propagates through a curved
spacetime, the anisotropy of the radiation pattern tends to be
somewhat reduced due to the phenomenon of {\it tails} --- essentially,
the backscatter of the radiation from spacetime curvature itself.
Both the redshifting and the anisotropy reduction reduce the recoil
relative to Fitchett's original analysis.

\begin{figure}[t]
\begin{center}
  \includegraphics[width=.45\textwidth]{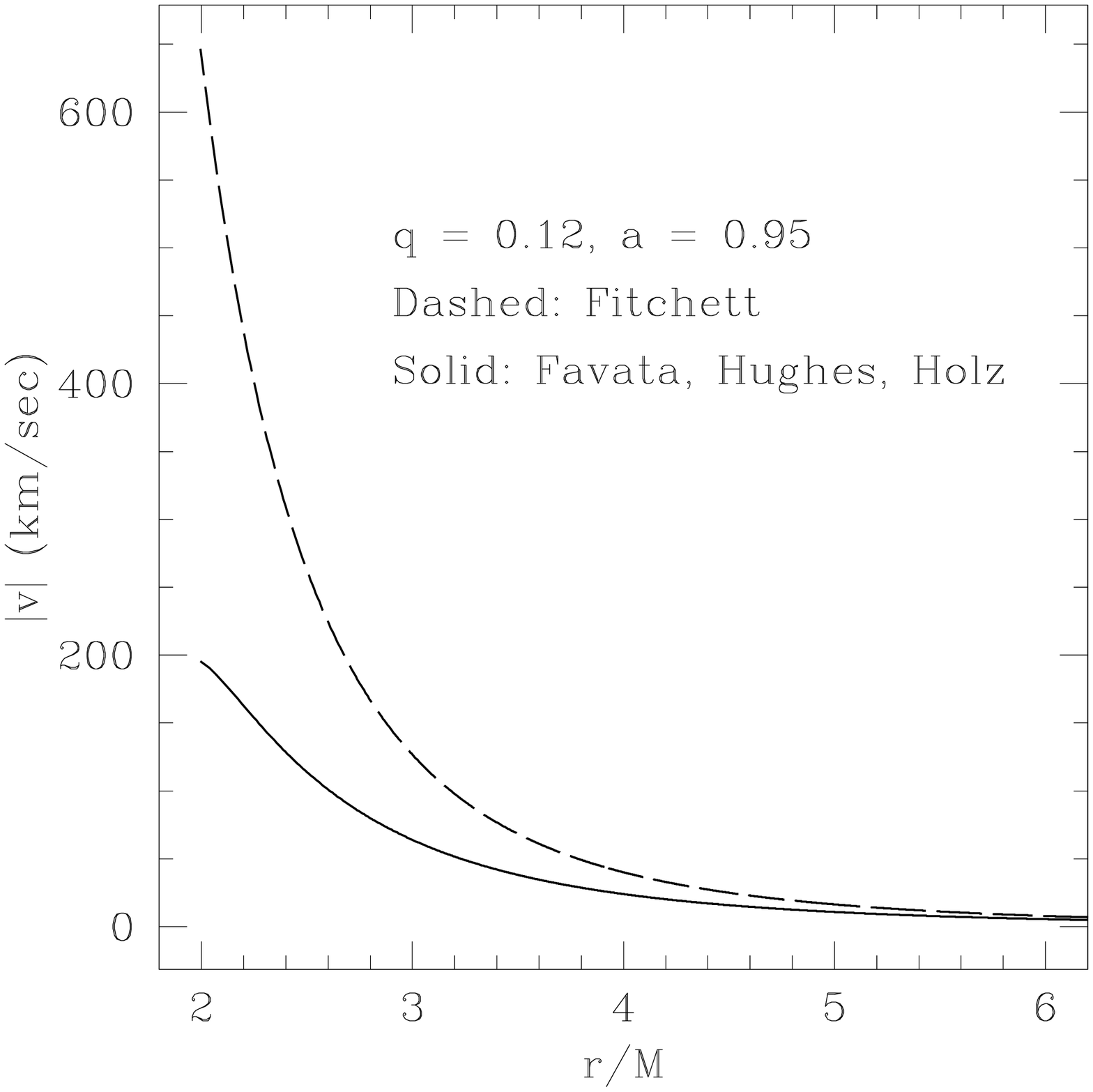}
  \includegraphics[width=.45\textwidth]{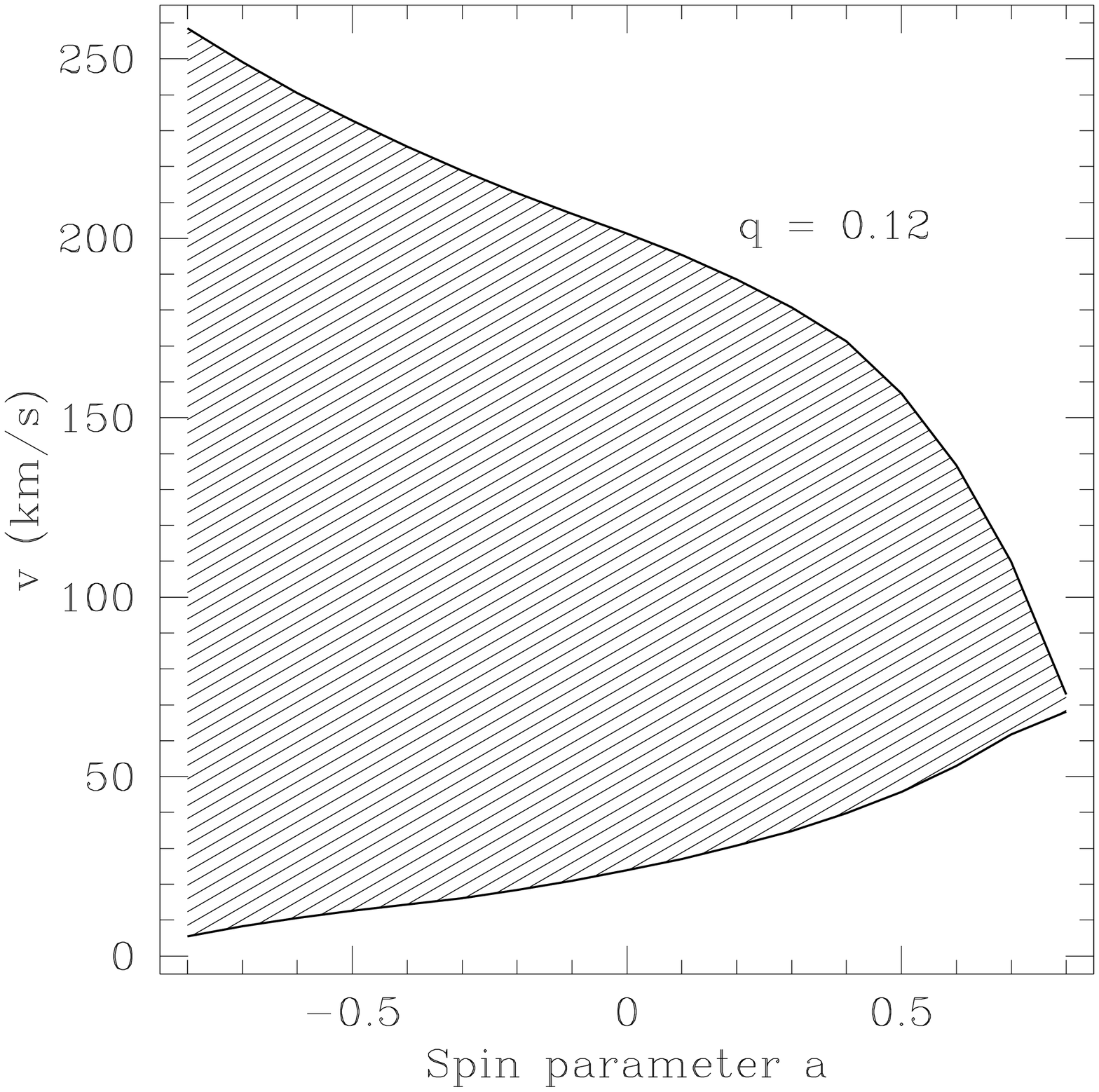}
\caption{Left: Fitchett's accumulated recoil and our results, versus
binary separation $r$.  This illustrates the importance of a proper
treatment of strong-field physics.  Right: Total recoil versus black
hole spin $a$.  The upper curve is our overestimate, the lower our
underestimate.  The span between them shows the importance of the
final plunge.}
\label{fig:results}
\vskip -3mm
\end{center}
\end{figure}

The right-hand panel of Fig.\ {\ref{fig:results}} summarizes our
results for the total recoil that can be expected in a merger; the
curves shown are for $q = 0.12$, but can be rescaled using $f(q)$
[Eq.\ (\ref{eq:vfitchett}) and subsequent text].  The range shown here
shows the uncertainty that results from our inability to model the
plunge very well.  It is largest for retrograde orbits ($a < 0$) of
rapidly rotating black holes because the transition to plunge occurs
at relatively large radius there --- the smaller body plunges quite a
distance before passing through the event horizon.  By contrast, for
prograde orbits ($a > 0$), the transition occurs at small radius, so
the plunge does not matter quite so much.  Our uncertainty is much
smaller for those cases.

A very notable feature of this plot is that the recoil, though
substantial, {\it never} exceeds about 500 km/sec, even when the mass
ratio is ``tuned'' to maximize the recoil.  On the other hand, it is
not difficult for the recoil to reach several tens of km/sec, even for
our underestimate.  Indeed, when we convolve our over- and
underestimates with a distribution of likely mass ratios and spin
values (cf.\ Ref.\ {\cite{mmfhh04}}) we find that recoils of several
tens of km/sec are quite easy to achieve; recoils $\sim 100$ km/sec
are likely for an interesting fraction of recoils; and recoils of
several hundred km/sec, though possible, are probably rather rare.

In the near future we hope to reduce our error bars, but for now we
understand the recoil with sufficient accuracy that these results can
be used for many astrophysical applications The most likely recoil
range --- several tens to a few hundred km/sec --- is particularly
interesting: While not large enough to eject black holes from massive
galaxies, kicks in this range can lead to ejection in dwarf galaxies
and dark matter halos, affect the nuclear density profiles of galaxies
with SMBHS, and influence the hierarchical growth of supermassive and
intermediate mass black holes
{\cite{mmfhh04,mq04,vhm03,gmh04,h04,yme04,kmq04}}.

It is a pleasure to thank Saul Teukolsky and Jerry Ostriker for
bringing this problem to our attention, and \'Eanna Flanagan, Avi
Loeb, David Merritt, Milo\v s Milosavljevi\'c, Martin Rees, Joseph
Silk, Alan Wiseman, and Yanqin Wu for many useful discussions.  SAH
would especially like to thank the organizers of this conference for a
fantastic meeting.  MF is supported by NSF grant PHY-0140209; SAH by
NSF grant PHY-0244424 and NASA grant NAG5-12906; and DEH by NSF grant
PHY-0114422.

%

\end{document}